\documentclass{article}

\usepackage{PRIMEarxiv}

\usepackage[utf8]{inputenc} % allow utf-8 input
\usepackage[T1]{fontenc}    % use 8-bit T1 fonts
\usepackage{hyperref}       % hyperlinks
\usepackage{url}            % simple URL typesetting
\usepackage{booktabs}       % professional-quality tables
\usepackage{amsfonts}       % blackboard math symbols
\usepackage{nicefrac}       % compact symbols for 1/2, etc.
\usepackage{microtype}      % microtypography
\usepackage{lipsum}
\usepackage{fancyhdr}       % header
\usepackage{graphicx}       % graphics

\usepackage[nolist]{acronym}
\usepackage{doi}

\title{Converging Safety and Security: IO-Link Wireless and OPC UA over 5G under prEN\,50742
\thanks{This is the author’s version of a paper that has been accepted for presentation at the 31st IEEE International Conference on
Emerging Technologies and Factory Automation (ETFA 2026), to be held in Västerås, Sweden, on September 08–11, 2026.}
}

\author{
 \href{https://orcid.org/0000-0002-5390-3946}{\includegraphics[scale=0.06]{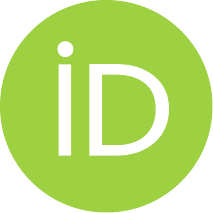}\hspace{1mm}Henry Beuster}\\
	Electrical Measurement Engineering\\
	Helmut-Schmidt-University\\
	Hamburg, Germany\\
	\texttt{henry.beuster@hsu-hh.de} \\
  \And
  \href{https://orcid.org/0000-0002-6882-1214}{\includegraphics[scale=0.06]{orcid.pdf}\hspace{1mm}Thomas Doebbert}\\
	Jungheinrich AG\\
	Norderstedt, Germany\\
	\texttt{thomas.doebbert@jungheinrich.de} \\
  \And
  \href{https://orcid.org/0009-0008-4470-2879}{\includegraphics[scale=0.06]{orcid.pdf}\hspace{1mm}Gerd Scholl}\\
	Electrical Measurement Engineering\\
	Helmut-Schmidt-University\\
	Hamburg, Germany\\
    \texttt{gerd.scholl@hsu-hh.de} \\
}

\begin{document}
\begin{acronym}
\acro{iolw}[IOLW]{IO-Link Wireless}
\acro{iolws}[IOLWS]{IO-Link Wireless Safety}
\acro{plc}[PLC]{programmable logic controller}
\acro{sfrt}[SFRT]{Safety Function Response Time}
\acro{iol}[IOL]{IO-Link}
\acro{revpi}[RevPi]{Revolution Pi}
\acro{vpn}[VPN]{virtual private network}
\acro{rssi}[RSSI]{Received Signal Strength Indicator}
\acro{iols}[IOLS]{IO-Link Safety}
\acro{wmaster}[W-Master]{Wireless-Master}
\acro{ds2ccp}[DS2CCP]{Digital Sensor-2-Cloud Campus Platform}
\acro{vr}[VR]{virtual reality}
\acro{estop}[e-stop]{emergency stop}
\acro{srsl}[SRSL]{Safety-Related Security Level}
\acro{opcua}[OPC\,UA]{OPC Unified Architecture}
\acro{embb}[eMBB]{Enhanced Mobile Broadband}
\end{acronym}
\maketitle

\begin{abstract}
The integration of wireless communication technologies in industrial automation offers greater flexibility, but also exposes safety systems to a broader threat vector. Emerging regulations, such as the draft standard prEN\,50742, mandate the convergence of functional safety and cybersecurity by requiring cryptographic security mechanisms directly in safety-critical communication. This paper presents an empirical evaluation of this safety-security convergence across a complete control chain, spanning from an IO-Link Wireless Safety device to a PLC via an OPC UA backbone. We measure the latencies and jitter of different Safety-Related Security Levels under prEN\,50742 over Ethernet, Wi-Fi\,6, and private 5G. Our results reveal that while cryptographic execution time is negligible, the resulting frame payload expansion severely restricts wireless fieldbus capacity, reducing the maximum number of devices per IO-Link Wireless track from 8 to 2. Furthermore, we demonstrate that, despite higher average latency, a private 5G provides sufficiently deterministic latency characteristics to preserve functional safety watchdog margins, unlike unlicensed Wi-Fi\,6.
\end{abstract}

% keywords can be removed

\section{Introduction}
The digital transformation of industrial automation is driving the integration of wireless networks across all levels, from field sensors to cloud systems. Architectures combining \ac{iolw} at the field level, \ac{opcua} as the interoperability backbone, and 5G networks as the transport medium enable highly flexible and scalable cyber-physical production systems.

Traditionally, functional safety protocols rely on the black-channel principle, using non-cryptographic checks (e.g., CRC, sequence numbers) to detect transmission errors. However, emerging regulations such as the draft standard prEN\,50742 \cite{pren50742} mandate the integration of cryptographic security mechanisms directly into the safety-critical data stream.

Although these mechanisms enhance security, their timing and packet size overheads might violate safety watchdogs and reduce availability. The empirical impact of different \acp{srsl} under prEN\,50742 on timing behavior over wireless links remains unquantified. To address safety-security convergence, a testbed bridging \ac{iolws} and \ac{opcua} Safety is implemented to measure the timing and capacity behavior over Ethernet, Wi-Fi\,6, and private 5G under different security configurations.

The remainder of this paper is organized as follows: Section~\ref{sec:stateofart} reviews the state of the art, Section~\ref{sec:methodology} presents the security-for-safety methodology, Section~\ref{sec:architecture} details the testbed architecture, Section~\ref{sec:evaluation} discusses the empirical results, and Section~\ref{sec:conclusion} concludes the paper.

\section{State of the Art}\label{sec:stateofart}
\subsection{Regulations and Standards}
The convergence of safety and security in industrial automation is driven by evolving European regulatory frameworks (Machinery Regulation (EU)\,2023/1230 and the Cyber Resilience Act (EU)\,2024/2847 (CRA)).

The Machinery Regulation introduces explicit requirements that connectivity must not lead to hazardous situations, particularly where external or remote connections can influence machine behavior. The draft standard prEN\,50742 provides the technical framework to implement these requirements by defining measures to prevent corruption of safety-related data and control functions.

In parallel, the CRA establishes baseline cybersecurity requirements for products, including secure design, vulnerability management, and lifecycle support, without explicitly addressing the interaction between cyber threats and functional safety. This is the focus of prEN\,50742, that can be interpreted as a safety-specific extension of general security requirements.

From a standardization perspective, prEN\,50742 complements and bridges established frameworks in both safety and security domains. Functional safety is traditionally governed by IEC\,61508, which defines the lifecycle-based approach to hazard analysis and risk reduction for electronic systems. IEC\,61784‑3 specifies functional safety communication. In parallel, IEC\,62443 provides a comprehensive framework for industrial security. prEN\,50742 integrates these perspectives by focusing on the impact of data integrity and system connectivity on safety functions.

\subsection{Wireless Automation}
Deploying safety-critical applications over wireless media requires stringent determinism and reliability \cite{wollschlaeger2017future}. At the field level, \ac{iolw} (IEC\,61139-3 \cite{iec61139-3}) achieves cable-grade reliability via adaptive frequency hopping and time-division scheduling in the 2.4\,GHz band \cite{doebbertStudySafeSecure2021a}. While IO-Link Safety (IEC\,61139-2) defines safety over wired links, the wireless extension is not yet standardized. We previously proposed a black-channel protocol architecture for \ac{iolws} \cite{doebbert2024diss} and demonstrated its feasibility with a roaming emergency-stop prototype \cite{beuster2024iolws-roaming}.

For backbone communication, \ac{opcua} provides transport-layer independence, making it well suited for Wi-Fi\,6 and private 5G. The comparative performance and co-existence of these two wireless technologies have been evaluated in various industrial IIoT settings, highlighting the trade-offs between Wi-Fi\,6's low average latency under light traffic and private 5G's bounded latency distributions and scalability under load \cite{maldonado2021comparing, segura2024empirical}. The \ac{opcua} Safety extension defines a black-channel approach up to SIL\,3 \cite{opcua-safety}, which can be enhanced via PubSub and Time-Sensitive Networking \cite{pfrommer2018opcua-tsn}. Previously, we determined the baseline latency by measuring the \ac{sfrt} of non-secured safety communication over private 5G and \ac{iolw} without \ac{opcua} \cite{beuster2024sfrt}.

Integrating cryptographic security directly into safety protocols is now mandated by prEN\,50742. Wieczorek and Schiller \cite{wieczorekSafetyAnalyseFuerSecuritygeschuetzte2020} analytically showed that cryptography alters traditional safety fault models and recommended black-channel integration. However, the empirical timing and capacity overhead of different \acp{srsl} over heterogeneous wireless networks remains unquantified, which this work addresses.

\section{Security-for-Safety Methodology}\label{sec:methodology}
\subsection{Safety-Related Security Levels}
prEN\,50742 introduces \acp{srsl} as a central concept for addressing cybersecurity threats to functional safety. They provide a graded classification of protection requirements that define the extent to which safety-related functions must be protected against corruption.

The primary objective is to ensure that the integrity of safety-related functions is maintained even in the presence of cybersecurity threats. Unlike conventional IT security levels, \acp{srsl} are derived from a safety perspective, based on the potential consequences of corruption rather than purely on confidentiality, integrity, and availability goals.
At the lowest level, \ac{srsl}\,0, no specific security measures are required from a safety perspective, due to isolation or lack of safety impact. Safety is ensured through inherent system design and isolation.
\ac{srsl}\,1 introduces basic protection against unintentional, accidental or low-effort interference. Measures focus on fundamental integrity checks and controlled access, ensuring that common operational errors do not compromise safety.
At \ac{srsl}\,2, protection addresses intentional manipulation by moderately capable attackers in connected environments. This requires stronger safeguards, including authenticated access, integrity protection for communication and data, and traceability of relevant changes.
\ac{srsl}\,3 represents high protection against sophisticated attackers in highly interconnected systems. It requires a defense-in-depth approach, including strong cryptographic mechanisms, strict access controls, and comprehensive monitoring and traceability.
An extended level, \ac{srsl}\,4, may be considered for systems with very high exposure and critical safety impact, requiring the highest degree of assurance and resilience against highly capable adversaries.
Overall, \acp{srsl} follow an impact-driven, safety-oriented approach, ensuring that the integrity of safety functions is maintained under accidental and malicious corruption scenarios.

\subsection{SRSL Configuration Mapping}
To evaluate the timing impact, the conceptual safety-related security levels are mapped to concrete configurations at the field and backbone levels, as detailed in Table~\ref{tab:srsl_mapping}.

\begin{table*}[t]
\caption{\ac{srsl} Configuration Mapping}
\label{tab:srsl_mapping}
\centering
\begin{tabular}{lp{4.0cm}p{5.0cm}p{4.0cm}}
\hline
\textbf{\ac{srsl}} & \textbf{IOLW} & \textbf{OPC UA} & \textbf{Security Target} \\
\hline
Plain & No Safety, No Cryptography & No Safety, No Cryptography & Baseline Reference \\
SRSL 0 & Safety, No Cryptography & Safety, No Cryptography & Safety baseline \\
SRSL 1 & Safety + 4-byte MAC & Safety + HMAC-SHA256 & Data origin authenticity and integrity \\
SRSL 2 & Safety + 4-byte MAC & Safety + Aes128Sha256RsaOaep & Backbone confidentiality and integrity \\
SRSL 3 & Safety + AES-CCM (128 bit) + 4-byte MAC &  Safety + Aes128Sha256RsaOaep & Full encryption security \\ \hline
\end{tabular}
\end{table*}

At the field level (\ac{iolw}) under Plain and \ac{srsl}\,0, no cryptography is used. For \ac{srsl}\,1 and \ac{srsl}\,2, authenticity and integrity are protected by appending a 4\,B truncated MAC (generated via \texttt{mbedtls\_poly1305}) \cite{doebbert2024diss}. For \ac{srsl}\,3, the payload is encrypted and authenticated using AES-CCM with a 128-bit key to prevent physical-layer eavesdropping.

The backbone between the Edge Gateway and the \ac{plc} uses standard \ac{opcua} security policies. Plain and \ac{srsl}\,0 use \texttt{None} security mode. \ac{srsl}\,1 uses the \texttt{Sign} mode with the \texttt{Aes128Sha256RsaOaep} policy, calculating an HMAC-SHA256 signature. \ac{srsl}\,2 and \ac{srsl}\,3 use the \texttt{SignAndEncrypt} mode with the same policy, applying symmetric AES-128-CBC encryption and asymmetric RSA-OAEP for key exchange.

\section{Testbed Architecture}\label{sec:architecture}
\subsection{System Overview}
\begin{figure*}[t]
    \centering
    \includegraphics[width=1\linewidth]{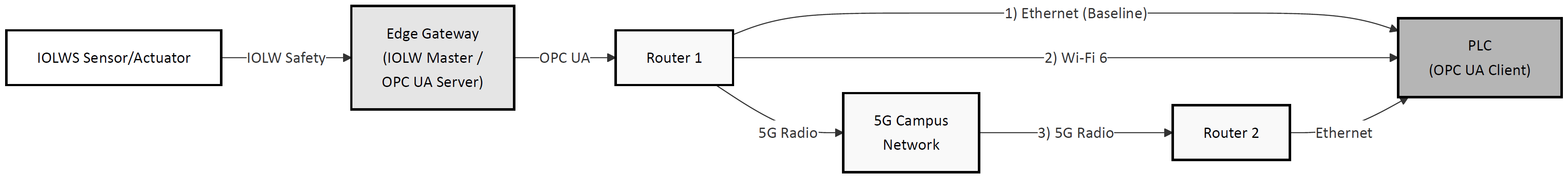}
    \caption{Converged safety-security system architecture and communication chain.}
    \label{fig:architecture_combined}
\end{figure*}

To evaluate the timing behavior, we implemented a demonstrator replicating a complete industrial control chain, from a field-level sensor node to a \ac{plc} in a backend environment, as depicted in Fig.~\ref{fig:architecture_combined}.

At the field level, a sensor/actuator node periodically samples process values at an application cycle time of 50\,ms. This is implemented on a TI CC2650 wireless MCU acting as an industrial \ac{iolws} device, transmitting via \ac{iolw} in the 2.4\,GHz band to the Edge Gateway. The Edge Gateway serves a dual function: it acts as the \ac{wmaster} and hosts an \ac{opcua} Server exposing the safety variables. It is implemented on a TI TMDSAM64GPEVM evaluation board running a Linux kernel.

Three communication channels are implemented between the Edge Gateway and the \ac{plc} (\ac{opcua} Client): (1) \textit{Wired Gigabit Ethernet (1000BASE-T)} as a low-latency, low-jitter baseline; (2) \textit{Unlicensed Wi-Fi\,6 (IEEE 802.11ax)} in the 5\,GHz band via an enterprise-grade access point; and (3) a \textit{Licensed Private 5G Network} operating in standalone SA mode (Band n78, 100\,MHz bandwidth from 3.7 to 3.8\,GHz) with a subcarrier spacing of 30\,kHz. The network was configured with a TDD frame pattern of 7:2 (DDDDDDDSUU) and ran on a eMBB network slice, utilizing an industrial 5G gateway.

\subsection{Field Level: IO-Link Wireless}
Field-level transmission uses \ac{iolw} with a 5\,ms cycle time, divided into three 1.66\,ms subcycles. To support safety, we use the \ac{iolws} protocol architecture \cite{doebbert2024diss}, treating the link as a black-channel. Safety data is encapsulated in a Wireless Safety Protocol Data Unit (W-SPDU) with a safety header (to detect packet loss, replication, or insertion) and a CRC-32. Cryptographic mechanisms (MAC/encryption) are layered based on the \ac{srsl} configuration (Table~\ref{tab:srsl_mapping}) using pre-shared keys to avoid handshake delays.

\subsection{Network Level: OPC UA}
The Edge Gateway maps the incoming \ac{iolws} frames directly to the \ac{opcua} information model and encapsulates them into an \ac{opcua} Safety SPDU for backbone transmission. The \ac{opcua} safety communication runs between the gateway and the \ac{plc}. Backbone transport security is enforced via standard \ac{opcua} Secure Channel policies according to the active \ac{srsl} configuration (Table~\ref{tab:srsl_mapping}).

\section{Experimental Evaluation}\label{sec:evaluation}
\subsection{Measurement Setup}
To validate the performance of the \ac{srsl} configurations specified in prEN\,50742, we constructed a measurement setup designed to isolate network transport latencies from cryptographic processing delays. The nodes are instrumented using hardware GPIO pins connected to a high-resolution digital oscilloscope to capture microsecond-level timings. We measure latencies along the communication chain, specifically at the over-the-air field level and the backbone network, and analyze statistical parameters over 10,000 communication cycles per scenario, with parallel network captures (Wireshark) monitoring packet overhead. The trials were conducted without RF isolation from the environment. Instead, measurements were performed in a typical office setting with active concurrent Wi-Fi networks, exposing the links to co-channel interference and background traffic, representing a realistic, non-shielded radio channel with the most pronounced impact expected on the unlicensed 5\,GHz Wi-Fi 6 backbone.

\subsection{Latency and Jitter Results}
To evaluate the impact of functional safety and cryptographic security layers on the timing behavior of the industrial communication chain, the measurements reflect the three physical media and five security configurations. Table~\ref{tab:iolw_measurements} and Table~\ref{tab:backbone_measurements} provide a comprehensive overview of the latency and standard deviation (jitter) profiles at the field and backbone levels.

\begin{table}[htbp]
\caption{Field-Level (IOLW) Protocol Overhead and Latency}
\label{tab:iolw_measurements}
\centering
\footnotesize
\setlength{\tabcolsep}{5pt}
\begin{tabular}{lccc}
\hline
\textbf{SRSL} & \begin{tabular}[b]{@{}c@{}}\textbf{Overhead}\\\textbf{down-/uplink [B]}\end{tabular} & \textbf{Latency $\pm$ $\sigma$ [ms]} & \textbf{Max Latency [ms]} \\ \hline
Plain   & 1/2  & $1.50 \pm 0.51$ & $2.70$ \\
SRSL 0  & 7/8  & $1.50 \pm 0.51$ & $2.70$ \\
SRSL 1  & 11/12 & $1.86 \pm 0.53$ & $2.90$ \\
SRSL 2  & 11/12 & $1.86 \pm 0.53$ & $2.90$ \\
SRSL 3  & 11/12 & $1.86 \pm 0.53$ & $2.90$ \\ \hline
\end{tabular}
\end{table}

\begin{table}[htbp]
\caption{Backbone (OPC UA) Protocol Overhead and Latency}
\label{tab:backbone_measurements}
\centering
\footnotesize
\setlength{\tabcolsep}{5pt}
\begin{tabular}{lccc}
\hline
\textbf{SRSL} & \textbf{Overhead [B]}& \textbf{Latency $\pm$ $\sigma$ [ms]} & \textbf{Max Latency [ms]} \\ \hline
\multicolumn{4}{c}{\textit{Local Wired Ethernet}} \\
Plain   & 132 & $1.78 \pm 0.51$ & $6.47$  \\
SRSL 0  & 128 & $4.29 \pm 1.43$ & $8.27$  \\
SRSL 1  & 148 & $4.24 \pm 1.46$ & $8.68$  \\
SRSL 2  & 155 & $4.12 \pm 1.53$ & $8.95$  \\
SRSL 3  & 155 & $4.12 \pm 1.61$ & $9.80$  \\ \hline
\multicolumn{4}{c}{\textit{Wi-Fi 6 Unlicensed Wireless}} \\
Plain   & 132 & $3.23 \pm 3.51$ & $76.76$ \\
SRSL 0  & 128 & $5.47 \pm 3.23$ & $85.85$ \\
SRSL 1  & 148 & $5.09 \pm 3.06$ & $89.54$ \\
SRSL 2  & 155 & $5.53 \pm 3.21$ & $72.86$ \\
SRSL 3  & 155 & $5.45 \pm 3.23$ & $76.81$ \\ \hline
\multicolumn{4}{c}{\textit{Private 5G}} \\
Plain   & 132 & $15.36 \pm 3.85$ & $29.53$ \\
SRSL 0  & 128 & $27.02 \pm 4.57$ & $52.37$ \\
SRSL 1  & 148 & $27.66 \pm 4.76$ & $52.15$ \\
SRSL 2  & 155 & $30.73 \pm 5.02$ & $61.39$ \\
SRSL 3  & 155 & $30.88 \pm 5.05$ & $59.44$ \\ \hline
\end{tabular}
\end{table}

\subsubsection{Field-Level Performance}
The baseline latency of \ac{iolw} is $1.50 \pm 0.51$\,ms for Plain and \ac{srsl}\,0. Introducing software cryptography at \ac{srsl}\,1--3 increases the average latency by a delta of $0.36$\,ms to $1.86 \pm 0.53$\,ms. Crucially, the standard deviation remains constant, confirming that software cryptography execution does not significantly degrade channel determinism.

However, the packet size expansion (from 1\,B overhead in Plain to 7\,B in \ac{srsl}\,0 and 11\,B in \ac{srsl}\,1--3) introduces a severe trade-off regarding the capacity of the wireless fieldbus operating at a 5\,ms cycle. In the non-safety configuration, up to 8 devices with 2\,B payload in the uplink can be scheduled per track. When transitioning to \ac{srsl}\,0, the safety header and CRC require double-slot devices with up to 15\,B payload in the uplink, reducing the number of devices per track to 4. At \ac{srsl}\,1--3, the introduction of the 4\,B MAC reduces the capacity to schedule only to 2 devices per track under this configuration, because of \ac{iolw}'s maximum downlink payload of 37\,B shared by all devices. Security regulations directly limit sensor density in this case.

\subsubsection{Backbone Latency}
The timing behavior over the backbone shows distinct patterns:
First, the transition from Plain to \ac{srsl}\,0 introduces a substantial latency penalty: $+2.51$\,ms on Ethernet ($1.78$ to $4.29$\,ms), $+2.24$\,ms on Wi-Fi\,6 ($3.23$ to $5.47$\,ms), and $+11.66$\,ms on 5G ($15.36$ to $27.02$\,ms). This additional latency is not caused by network delays but is instead a direct consequence of the safety-layer protocol, which requires two-way transactions (request/response) to verify the channel status. This mechanism doubles the transmission delay per cycle, making the safety function sensitive to the physical layer's one-way delay. At the same time, it slightly reduces the protocol overhead from 132\,B to 128\,B due to the changed communication pattern.

Second, the addition of backbone transport security has a minimal impact. Comparing \ac{srsl}\,0 to \ac{srsl}\,3, the latency remains almost identical on Ethernet and Wi-Fi\,6, and only increases by $+3.86$\,ms on the 5G network, which is well within the 5G jitter profile. This negligible timing impact is observed despite the backbone packet overhead increasing from 128\,B in \ac{srsl}\,0 to 155\,B in \ac{srsl}\,3 (Table~\ref{tab:backbone_measurements}).

Finally, comparing Wi-Fi\,6 and the private 5G network highlights a clear trade-off. Wi-Fi\,6 achieves lower average latency ($5.45$\,ms at \ac{srsl}\,3) but suffers from outliers exceeding 50\,ms. In contrast, the private 5G network exhibits higher average latency ($30.88$\,ms) but provides a predictable, bounded latency distribution without high-latency spikes (maximum latency of 59.44 ms). For safety communication, 5G's bounded worst-case latency allows engineers to configure a tighter and more reliable watchdog time, preventing spurious shutdowns.

\section{Conclusion}\label{sec:conclusion}
This paper presented an empirical evaluation of safety-security convergence in wireless automation networks, measuring the impact of different \acp{srsl} applying prEN\,50742.

The experimental results show that the main impact of cryptographic security is not the computational processing delay, which is negligible, but rather the expansion of the frame payload size. For fieldbus protocols such as \ac{iolw}, this payload inflation directly restricts device capacity per track, reducing the maximum supportable devices from 8 to 4, and ultimately to 2 devices for the short 5\,ms cycle time. Furthermore, a private 5G network might be preferable to Wi-Fi\,6 for safety functions because, despite Wi-Fi\,6’s lower average latency, 5G provides the physical-layer determinism and bounded worst-case latency necessary to prevent safety watchdog violations and avoid unnecessary system shutdowns in noisy industrial environments.

Future work will focus on testing the timing resilience of this converged architecture under high network loads, background traffic and multiple devices. Additionally, we plan to evaluate different OPC\,UA security configurations and the latency behavior of OPC\,UA PubSub implementations over wireless links, as well as to analyze different 5G network slice configurations.

\section*{Acknowledgment}
The authors would like to thank Telekom Deutschland GmbH and Ericsson GmbH for their continuous support and valuable cooperation throughout this work.

\section*{Funding}
This research paper out of the project “Digital Sensor-2-Cloud Campus Platform” (DS2CCP, \href{https://dtecbw.de/home/forschung/hsu/projekt-ds2ccp}{https://dtecbw.de/home/forschung/hsu/projekt-ds2ccp})) is funded by dtec.bw – Digitalization and Technology Research Center of the Bundeswehr. dtec.bw is funded by the European Union – NextGenerationEU.

%Bibliography
\bibliographystyle{unsrt}  
\bibliography{literature}

\end{document}